\documentclass[a4paper,fleqn]{cas-dc}

\usepackage[numbers,sort&compress]{natbib}

\def\tsc#1{\csdef{#1}{\textsc{\lowercase{#1}}\xspace}}
\tsc{WGM}
\tsc{QE}
\tsc{EP}
\tsc{PMS}
\tsc{BEC}
\tsc{DE}

\begin{document}
\newcommand{\angstrom}{\mbox{\normalfont\AA}}
\let\WriteBookmarks\relax
\def\floatpagepagefraction{1}
\def\textpagefraction{.001}

\title [mode = title]{Electronic and magnetic properties of graphene quantum dots with two charged vacancies}                      



\author{E. Bulut Kul}[
role=,orcid=0000-0003-2392-4313
]
\cormark[1]
\ead{bulutkul@iyte.edu.tr}


\address{Department of Physics, Izmir Institute of Technology, IZTECH, TR35430, Izmir, Turkey}

\author{M. Polat}[
role=Researcher,orcid=0000-0002-1372-1945
]

\author{A. D. G\"u\c{c}l\"u}[%
role=Co-ordinator,orcid=0000-0002-4351-7216
]

\credit{Data curation, Writing - Original draft preparation}

\cortext[cor1]{Corresponding author}
%

\begin{abstract}
Electronic and magnetic properties of a system of two charged
vacancies in hexagonal shaped graphene quantum dots are investigated
using a mean-field Hubbard model as a function of the Coulomb
potential strength $\beta$ of the charge impurities and the distance
$R$ between them. For $\beta=0$, the magnetic properties of the
vacancies are dictated by Lieb's rules where the opposite (same)
sublattice vacancies are coupled antiferromagnetically
(ferromagnetically) and exhibit Fermi oscillations. Here, we
demonstrate the emergence of a non-magnetic regime within the
subcritical region: as the Coulomb potential strength is increased to
$\beta \sim 0.1$, before reaching the frustrated atomic collapse
regime, the magnetization is strongly suppressed and the ground state
total spin projection is given by $S_z=0$ both for opposite and same sublattice
vacancy configurations. When long-range electron-electron interactions
are included within extended mean-field Hubbard model, the critical
value for the frustrated collapse increases from $\beta_{cf} \sim
0.28$ to $\beta_{cf} \sim 0.36$ for $R < 27 \ \angstrom$.

%
\end{abstract}

\begin{graphicalabstract}
\end{graphicalabstract}

\begin{highlights}
\item Investigation of electron-electron interaction effects in frustrated collapse in graphene systems
\item Emergence of non-magnetic regime within the subcritical collapse region.
\item Renormalization of frustrated critical coupling constant $\beta_{cf}$.
\end{highlights}


\maketitle

\section{\label{sec:level1}Introduction}
Recent advances at the atomic scale control of graphene through
vacancies \cite{Ugeda2010,Nair2012,Nair2013,Zhang2016}, charged
impurities \cite{Yan2011,Wang2015} and
adatoms \cite{Esquinazi2002,Esquinazi2003,Elias2009,Castro2009,McCreary2012} open up
possibilities for tailoring graphene's electronic and magnetic
properties \cite{Pereira2006,Palacios2009,CastroNeto2009,Yazyev2010,Mucciolo2010,Balakrishnan2014}
for future spintronic and computing applications, as well as for
investigating relativistic quantum effects such as atomic
collapse \cite{Wang2015,Levitov2007,Shytov2009,Gorbar2013,Valenzuela2016,Moldovan2017,Lu2019}. While
pure graphene is not expected to be magnetic, breaking of the
sublattice symmetry of the honeycomb lattice through atomic defects is
expected to exhibit local magnetization as predicted by theoretical
calculations \cite{Pereira2006,Pereira2008,Yazyev2007,Sevincli2008,Leconte2011,Zou2012,Pujari2009}. This local magnetization was recently observed experimentally using scanning tunneling
microscopy (STM) around hydrogen adatoms \cite{Herrero2016} and single
atomic vacancies \cite{Zhang2016}.

On the other hand, Mao et al. \cite{Mao2016} have shown that carbon
vacancies in graphene can host a stable positive effective charge $Z$
which can be gradually increased by applying STM voltage pulses. This
tunability of the coupling constant $\beta=Z \alpha_g$, where $\alpha_g=2.2/ \kappa$ is the effective fine-structure constant and $\kappa$ is the dielectric
constant, allows the observation of the system to
undergo a transition from subcritical to supercritical regime where
the 1S-like state dives into Dirac continuum, forming quasi-bound
states and mimicking the atomic collapse expected to occur in
ultra-heavy
nuclei \cite{Pomeranchuk1945,Werner1958,Pieper1969,Zeldovich1972,Reinhardt1977,Soff1985}
with $Z \sim 172$ \cite{Soff1974} which do not exists in
nature. Theoretically predicted by Pereira et al. \cite{Pereira2007},
the atomic collapse in graphene was first successfully observed through clusters of charged calcium dimers \cite{Wang2015}. On the other hand,
when two or more impurities with identical charges are present a
frustrated supercritical regime occurs at a distance dependent
critical value $\beta_{cf}$ which is lower than the critical value $\beta_{c}=0.5$
for a single charge impurity \cite{Wang2020,Pottelberge2019,Lu2019}.

An open question that we address in this work is, how do charged
vacancies magnetically couple to each other as a function of $\beta$.
For $\beta=0$, a theorem due to Lieb for bipartite Hubbard systems
predicts \cite{Lieb1989} that the local magnetic moments formed around
the vacancies should couple to each other ferromagnetically or
antiferromagnetically over large distances depending on whether they
lie on the same or opposite sublattices. Moreover, as the system is
reminiscent of Ruderman-Kittel-Kasuya-Yoshida (RKKY) model, one
expects to observe oscillations of magnetic coupling if the vacancies
are along the zigzag directions as opposed to a smooth decrease along
the armchair directions \cite{Guclu2015}. On the other hand, as $\beta$
is increased, Lieb's theorem does not apply anymore, the local
magnetization around vacancies is suppressed and one expects the
magnetic coupling between the two local moments to be severely
distorted.  

In this work, we consider a finite size graphene quantum dot
(GQD) \cite{Guclu2015,Zarenia2011,Sheng2012,Korhan2017,Guclu2018,GrapheneQuantumDots2014,Szaowski2013,Wang2008}
with hexagonal armchair edges to investigate the magnetic coupling
properties between the charged vacancies. The armchair edges make the
system free of additional edge state effects. Moreover, the critical
$\beta$ value for which the 1S state crosses the Dirac point is known
to be independent of the size of the quantum dot both within effective
mass approximation and mean-field Hubbard models \cite{Polat2020,
  VanPottelberge2017}. Thus, the hexagonal GQD system provides us with
a practical way to understand bulk properties as well. Here, we
perform mean-field Hubbard calculations to show that the magnetization
of the vacancies is strongly suppressed for $\beta>0.1$ which is in
the subcritical regime, i.e.,  lower than the frustrated critical value
$\beta_{cf} \sim 0.28$ for the range of $R$ studied here. As a result,
the ground state total spin projection of the double vacancy system reduces to
$S_z=0$ for both opposite (AB) and same (AA) sublattice
configurations. When we include long-range electron-electron
interactions within extended MFH approximation, $ \beta_{cf} $ is
renormalized from 0.28 to 0.36 by suppression of overscreening \cite{Pereira2008,Levitov2008,Kotov2012}. We also investigated the
effect of second nearest neighbor hopping $t_{nnn}$. For
$t_{nnn}=0.2$ eV, we found that Lieb's predictions for magnetization
of same sublattice vacancy system is violated even for $\beta=0$.

\section{\label{sec:level2}Model and Method}

We use a one-band mean-field Hubbard (MFH) model where the single
electrons states are written as a linear combination of $p_z$ orbitals
on every carbon atom since the sigma orbitals are considered to be
mainly responsible for mechanical stability of graphene. Including
long range interactions, the extended mean-field Hubbard Hamiltonian
can be written as
\begin{align}
H_{MFH} & =\sum_{ij \sigma}t_{ij}(c^{\dagger}_{i \sigma}c_{j
	\sigma}+h.c.)\nonumber \\
& + U\sum_{i \sigma}(\langle n_{i \bar{\sigma}}\rangle -
\frac{1}{2} )n_{i \sigma}\nonumber \\
 & + \sum_{ij}V_{ij}(\langle n_{
	j}\rangle - 1)n_{i} \nonumber \\
& +\sum_{i\sigma}
V_{C}({\bf r}_i)c^{\dagger}_{i\sigma}c_{i\sigma}
\end{align}

where the first term represents the tight-binding Hamiltonian and
\(\ t_{ij} \)'s are the hopping parameters given by \(\ t_{nn}=-2.8 \)
eV for nearest neighbors and \(\ t_{nnn}=-0.1 \) eV for next
nearest-neighbors \cite{CastroNeto2009}. Additionally, in this work
we considered \(\ t_{nnn}=0 \) eV and \(\ t_{nnn}=-0.2 \) eV to
investigate the stability of Lieb's theorem against \(\ t_{nnn}
\). The \(\ c^{\dagger}_{i \sigma}\) and \(\ c_{i \sigma}\) are
creation and annihilation operators for an electron at the i-th
orbital with spin \(\ \sigma\), respectively. Expectation value of
electron densities are represented by \(\ \langle n_{i \sigma} \rangle
\). The second term represents on-site Coulomb interactions. We take
on-site interaction parameter as \(\ U = 16.522/ \kappa \) eV, with
effective dielectric constant \(\ \kappa = 6 \) to take into account screening effects due to substrate \cite{Ando2006}. The
third term stands for long-range Coulomb interaction. Interaction
parameters \(\ V_{ij} = 8.64 / \kappa \) and \(\ V_{ij} = 5.33 /
\kappa \) for the first and next nearest neighbors respectively,
numerically calculated using Slater \(\ p_{z} \) orbitals \cite{Guclu2010}. Beyond second nearest neighbors, interactions are
calculated assuming point charges. Finally, the last term represents
the Coulomb potential due to vacancy charges located at ${\bf R}_1$ and
${\bf R}_2$, expressed as
\begin{equation}
V_{C}({\bf r}_i)=
-\hbar v_F \beta
\left(  \frac{1}{\vert {\bf r}_i-{\bf R}_1 \vert} 
+\frac{1}{\vert{\bf r}_i-{\bf R}_2 \vert} \right) 
\end{equation}
where $v_F = 3at/2 \ (\thicksim10^6 m/s) $ is the Fermi velocity. The dimensionless Coulomb potential strength 
$\beta$ can be tuned as discussed above. In this work, we assume that the charged impurities cause ideal vacancies in the honeycomb lattice where relaxation and bond reconstruction effects are neglected.

The hexagonal armchair quantum dot system that we consider in this
work consists of 5512 atoms for MFH calculations and up to 10806
atoms for TB calculations. A critical step in the numerical
calculations is the initial guess state used for the self-consistent
diagonalization of the MFH Hamiltonian. We have used various initial
guess spin states to ensure to the convergence to lowest possible ground
states consistent with the two competing total spin projections $S_z=1$
(ferromagnetic coupling) and $S_z=0$ (antiferromagnetic coupling).

\begin{figure}
	\includegraphics[scale=0.586]{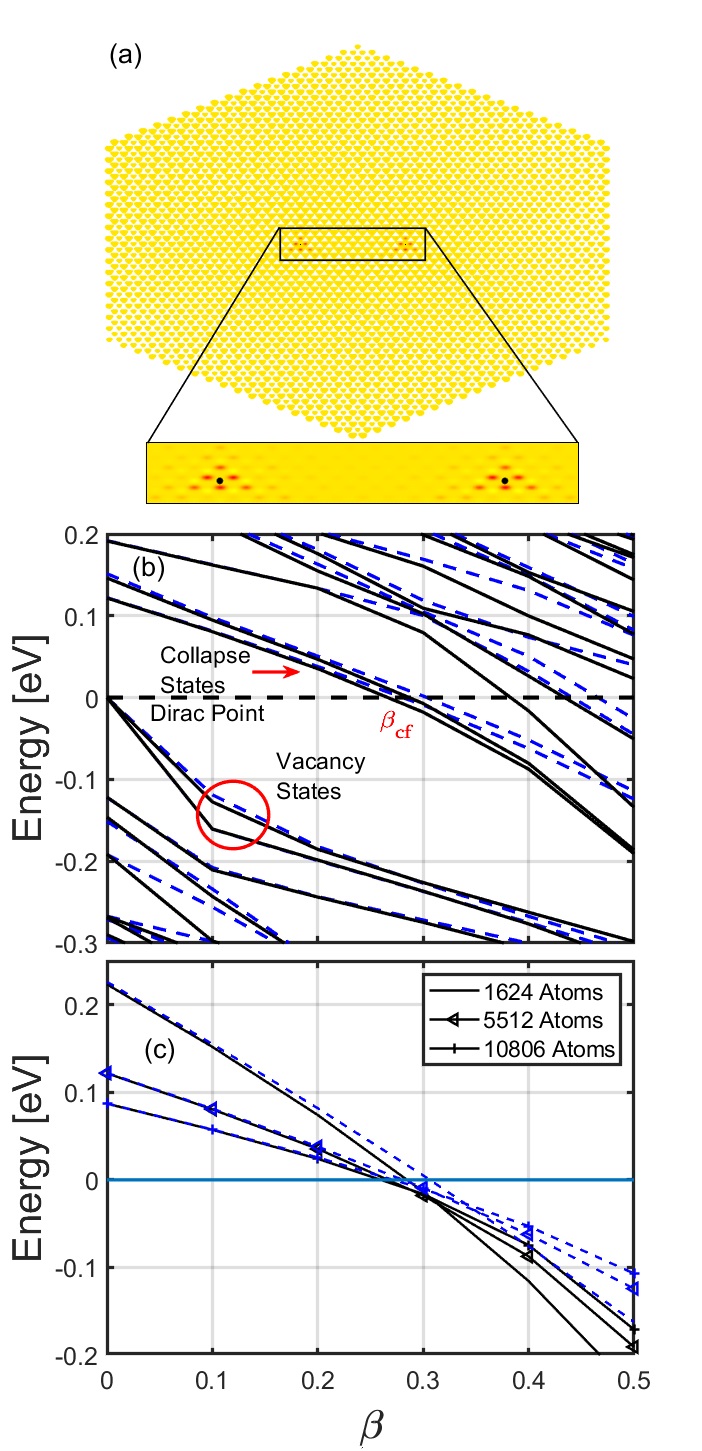}
	\caption{(a) Cross section image of electron density for a
          hexagonal armchair GQD, 5512 atoms, with two vacancies for
          AA case. Inter-vacancy distance is set to $ R=11b $ where $b$
          is second nearest-neighbor distance. Black dots represent
          vacancy positions. (b) TB energy spectrum versus
          $\beta$ and (c) GQD size comparison for $
          R/b=3 $ (black-solid lines) and $ R/b=9 $ (blue-dashed
          lines).}
	\label{fig:Fig1}
\end{figure}

\clearpage
\section{\label{sec:level3}Results and Discussion}
\subsection{Tight-binding results}	
As mentioned above, we consider AA and AB configurations for
vacancies located along the zigzag direction and separated by a
distance $R$, as shown in Fig. \ref{fig:Fig1}a for the AA
configuration. The midpoint between the vacancies is chosen to be the
center of the dot to minimize edge effects. Fig. \ref{fig:Fig1}b shows
a typical tight-binding (TB) energy spectrum as a function of $\beta$
in the vicinity of the Dirac point, obtained for the AA configuration
with $R/b=3$ (solid lines) and $R/b=9$ (dashed lines) where $b=2.46$
{\angstrom} is the second nearest neighbor distance. As expected,
there are two sets of vacancy states and collapse states corresponds to
to the bonding and anti-bonding states \cite{Wang2020,Pottelberge2019} of two charged vacancies. Collapsing states cross the Dirac level at the
critical value $\beta_{cf} \sim 0.28$ indicating the lower limit for
the frustrated supercritical regime before the system enters the
molecular collapse regime at
$\beta_c=0.5$ \cite{Wang2020,Pottelberge2019,Lu2019}.
The lower value of $\beta_{cf}=0.28$ for the double impurity system is
expected since the Coulomb potential due to each impurity feed each
other, accelerating the collapse. This effect is expected to vanish
for large distances $R$. For the range of $R$ values studies in this
work, $\beta_{cf}$ is nearly constant. More importantly, $\beta_{cf}$
is also found to be largely independent of finite size effects for
dots larger than few thousands atoms, consistent with single charged
impurity results \cite{Polat2020} as seen in Fig.\ref{fig:Fig1}c,
provided $R$ is smaller than the dot diameter.  We also note that,
increasing $\beta$ lifts the degeneracy of the vacancy states
initially. The energy gap between the vacancy states increases up to
$\beta \sim 0.1$ but, starts decreasing again as $\beta$ is increased
further, pointing to a decoupling of bonding and anti-bonding vacancy
states at large $\beta$ values. This observation have important
consequences for the understanding of mean-field Hubbard results
discussed below.

\subsection{Mean-Field Hubbard results for bare vacancies}

\begin{figure}[b]
	\centering
	\includegraphics[scale=0.3]{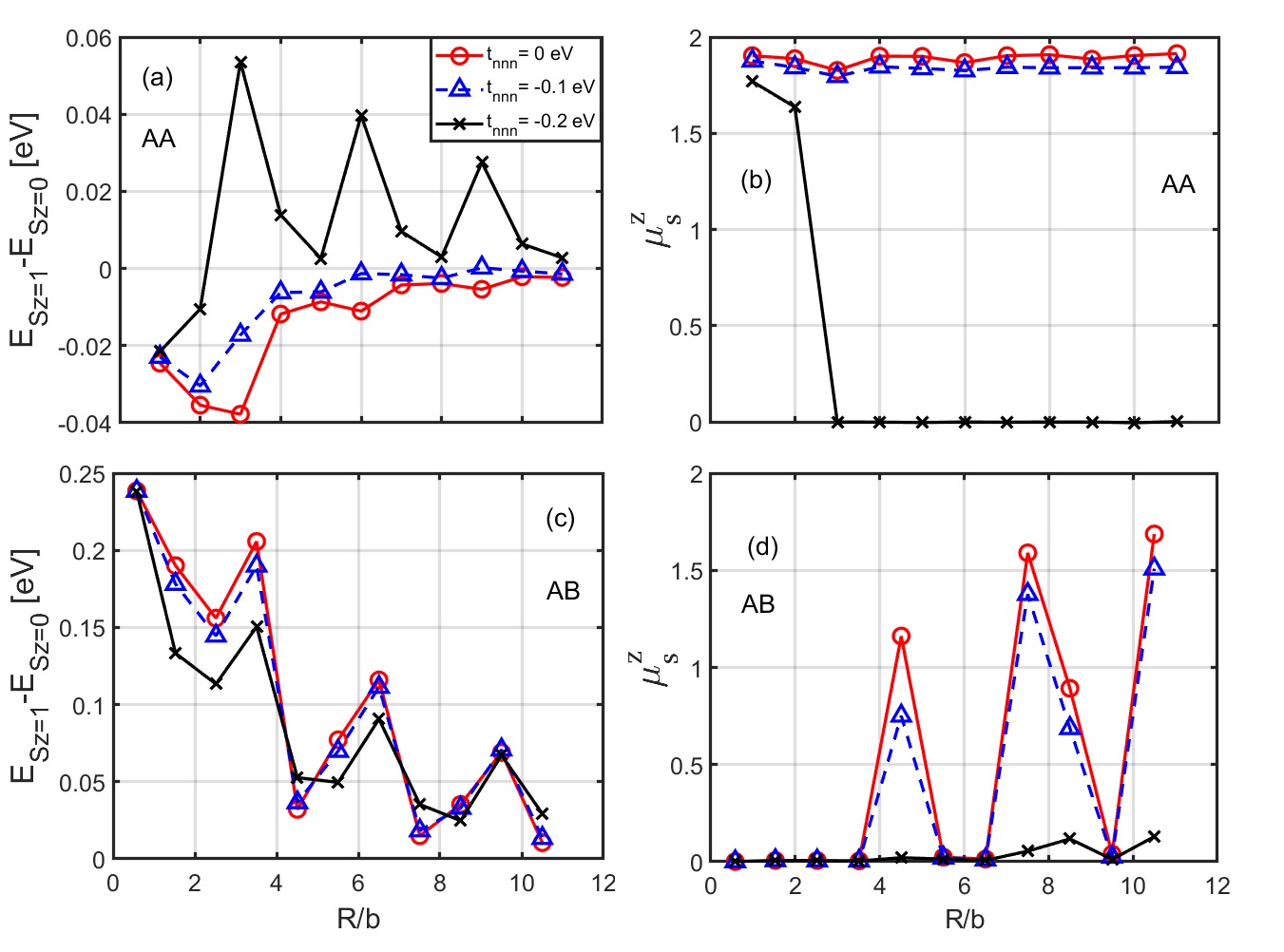}
	\caption{(a,c) Ground state energy differences $
          E_{S_z=1}-E_{S_z=0} $ for AA and AB cases and $ \beta=0$,
          (b,d) corresponding staggered magnetisms versus $ R/b$.
          Results are obtained using MFH method for hexagonal armchair
          GQD with 5512 atoms for different second nearest neighbor
          hopping parameters $t_{nnn}$.}
	\label{fig:Fig2} 
\end{figure}

In order to understand magnetic properties, we first focus on
$\beta=0$ and examine the stability of magnetic coupling between the
vacancies. Figure \ref{fig:Fig2}a and \ref{fig:Fig2}c show the spin
gap \(\ E_{S_z=1} - E_{S_z=0} \) as a function of distance $R$ for AA
and AB configurations respectively, obtained using MFH (without long
range interactions $V_{ij}$) for different second nearest neighbor
hopping parameters $t_{nnn}$. For $t_{nnn}=0$, ferromagnetic ($S_z=1$)
ground state for AA configuration and antiferromagnetic ($S_z=0$)
ground state for AB configuration are obtained as expected. Moreover,
the observed distance dependent oscillations are reminiscent of RKKY model for graphene along zigzag
direction \cite{Blackschaffer2010,Guclu2015}, assuming
$E_{S_z=1}-E_{S_z=0}$ is proportional to the effective magnetic
coupling parameter $J$ in the RKKY model. Here, however, the spins are
localized on three atoms neighboring each vacancy unlike in the RKKY
model. We have also investigated (not shown) the behavior of
$E_{S_z=1}-E_{S_z=0}$ as a function of distance along the armchair
direction and found a smooth decrease without oscillations, again
consistent with RKKY results. On the other hand, Fig. \ref{fig:Fig2}a
shows that the magnetic coupling between the bare vacancies is
strongly affected by $t_{nnn}$. For $t_{nnn}=-0.1$ eV, the value
usually accepted for graphene systems, the oscillations lose their
characteristic period of $3a$, where $a=2.46 \ \angstrom$ is the lattice constant of graphene. Moreover, for  $t_{nnn}=-0.2$ eV, the
ground state total spin projection becomes $S_z=0$, and the staggered magnetization
defined as $(-1)^x(n_{i \downarrow}-n_{i\uparrow })/2$ where x is even
for A and odd for B sublattice sites, is completely suppressed as
shown in Fig. \ref{fig:Fig2}b. The losing of staggered magnetization is
also observed for the AB configuration as shown in
Fig. \ref{fig:Fig2}d. These results shown that magnetic properties of
the double vacancy system are sensitive to $t_{nnn}$.  In the remaining
of this work, $t_{nnn}$ will be set to zero.

\subsection{Mean-Field Hubbard results for charged vacancies}

\begin{figure}
	\centering
	\includegraphics[scale=0.3]{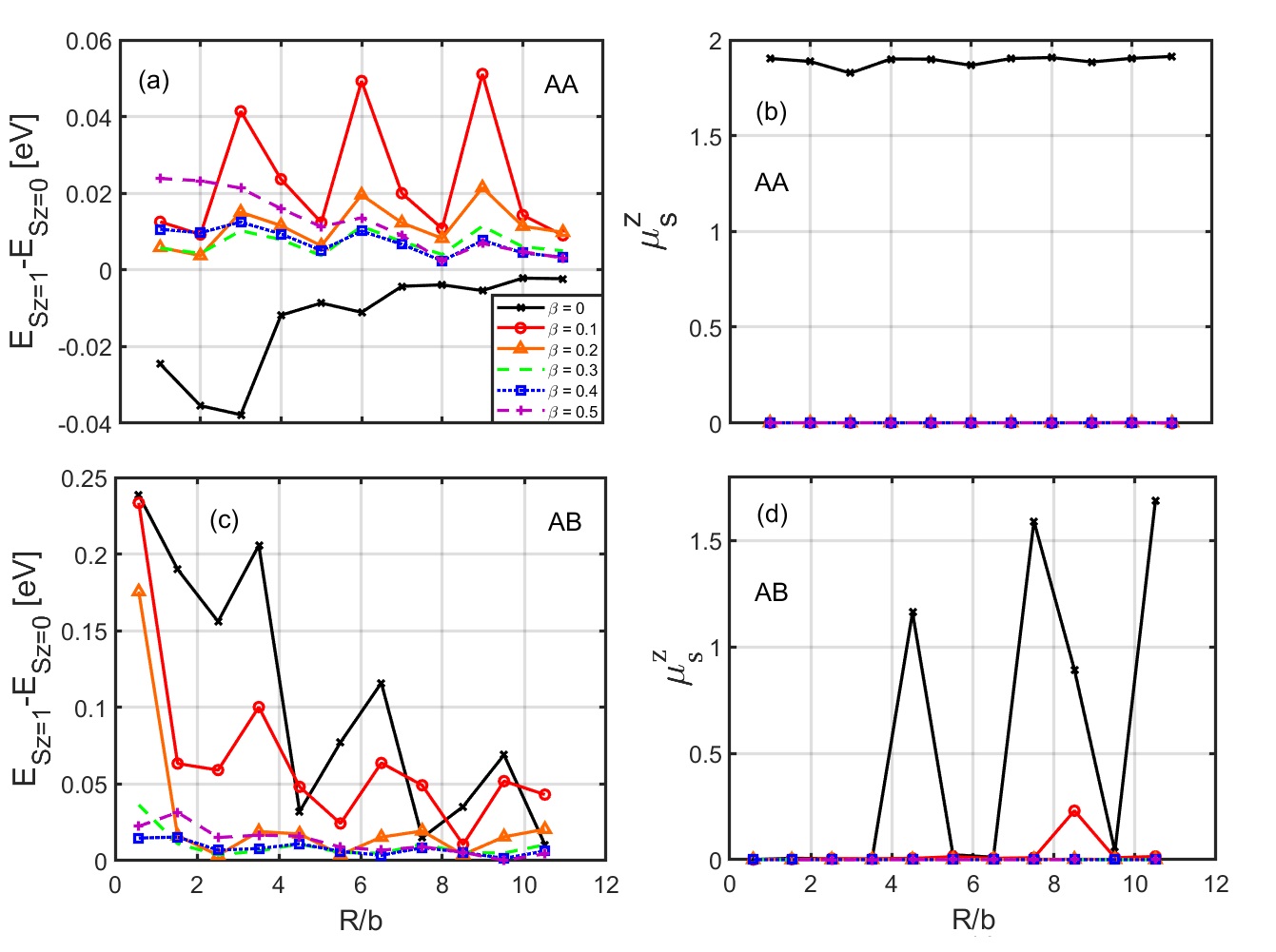}
	\caption{(a,c) Ground state energy differences $
          E_{S_z=1}-E_{S_z=0} $ and (b,d) corresponding staggered
          magnetisms versus $ R/b$ for different $ \beta$ values
          obtained using MFH method for hexagonal armchair GQD with
          5512 atoms.}
	\label{fig:Fig3}  
\end{figure}

We now investigate the effect of the Coulomb coupling strength
$\beta$. Figures \ref{fig:Fig3}a and \ref{fig:Fig3}c show
$E_{S_z=1}-E_{S_z=0}$ as a function of $R/b$ for different values of
$\beta$, for the AA and AB configurations respectively obtained using
MFH calculations excluding long-range electron-electron
interactions. Even at low values of $\beta=0.1$, $S_z=0$ becomes the
ground state for AA configurations, and staggered magnetization is
quenched (see Fig. \ref{fig:Fig3}b). A similar quenching of staggered
magnetization is also observed for the AB configuration. As $\beta$ is
increased further, spin gaps gradually approach zero for both AA and
AB configurations.

\begin{figure}
	\centering
	\includegraphics[scale=0.36]{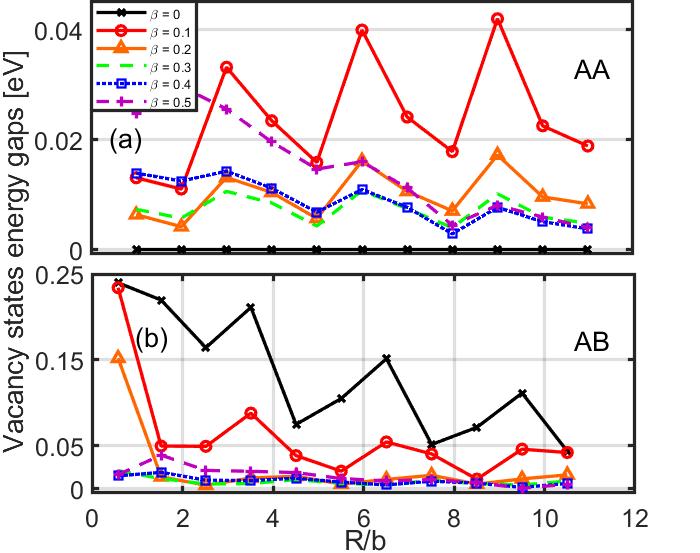}
	\caption{TB vacancy states gap for (a) AA and (b) AB cases along
          zigzag direction versus \(\ R/b \) for different \(\ \beta\)
          values.}
	\label{fig:Fig4}  
\end{figure}

In order to understand the behaviour of the spin gap shown in
Fig. \ref{fig:Fig3} further, we plot the
tight-binding energy differences between the two vacancy states as a
function of $R/b$ for different $\beta$ values as shown in Fig. \ref{fig:Fig4}. Clearly, the tight-binding energy gaps show qualitatively similar features compared to the spin gaps shown in Fig. \ref{fig:Fig3}a and c, except for the AA configuration at $\beta=0$. Indeed the vacancy states are degenerate in this latter case and the spin gap is dominated by electron-electron interactions, leading to an effective ferromagnetic interaction. In other cases, the degeneracy is lifted and the spin gaps are mainly dictated by tight-binding kinetic energies.

\subsection{Extended Mean-Field Hubbard Method calculations}

\begin{figure}
	\centering
	\includegraphics[scale=0.3]{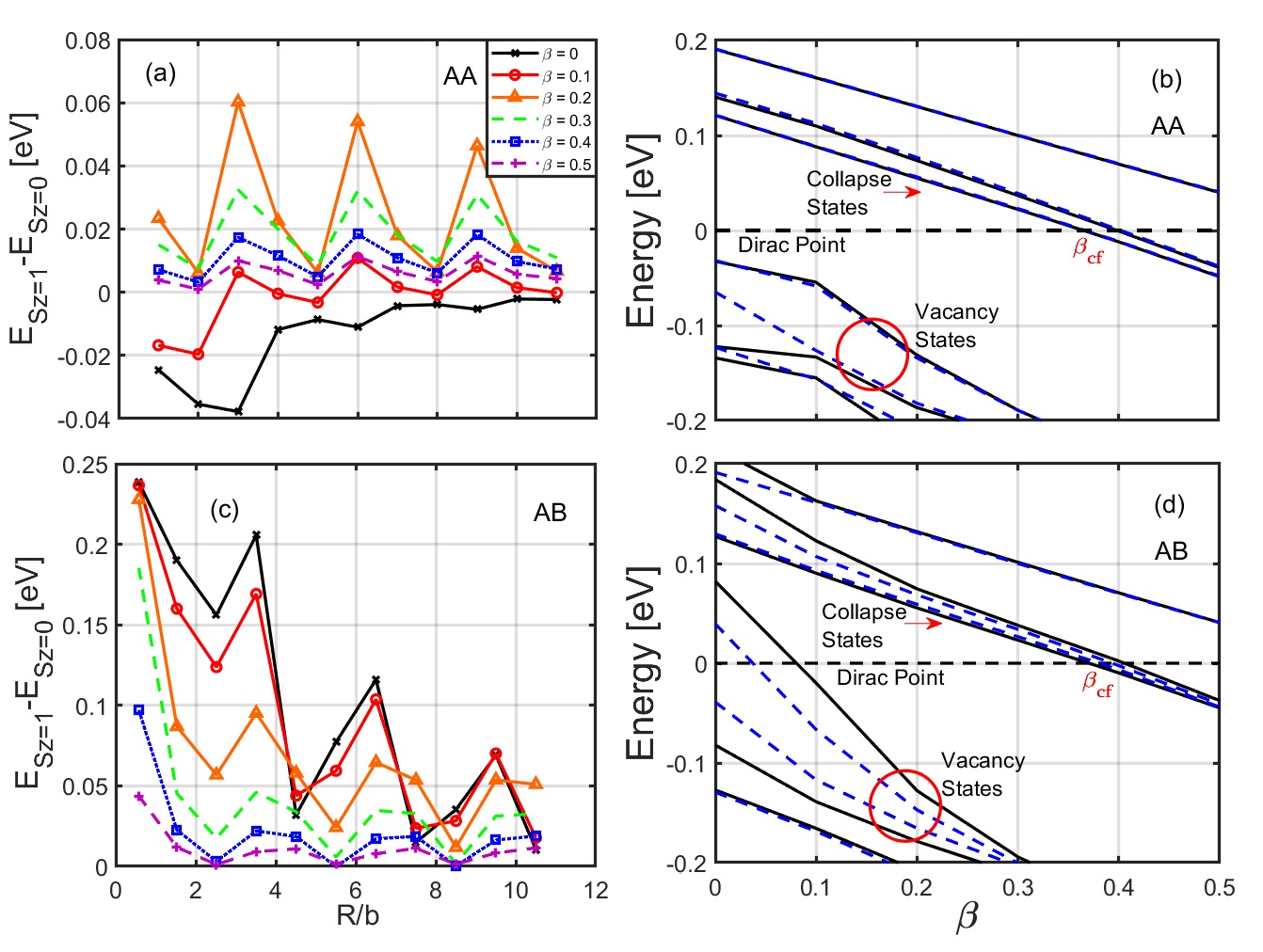}       
	\caption{(a,c) Ground state energy differences
          \(\ E_{S_z=1}-E_{S_z=0} \) versus $R/b$ for different
          $\beta$ values, and (b,d) energy spectra versus $\beta$ for
          $R/b=3$ (black-solid lines) and $R/b=9$ (blue-dashed
          lines). Results are obtained using extended MFH method for GQD's
          with 5512 atoms.}
	\label{fig:Fig5}     
\end{figure}

We now investigate the effects of long-range electron-electron
interaction terms $V_{ij}$. Since charged impurities causes the charge
distribution to be inhomogeneous, long-range electron interactions can
be expected to play an important role. Figure \ref{fig:Fig5} shows the
spin gaps \(\ E_{S_z=1} - E_{S_z=0} \) and energy spectra for for spin
up electrons obtained using the extended mean-field Hubbard
model. Although Figs. \ref{fig:Fig5}a,c are qualitatively similar to
Figs. \ref{fig:Fig3}a,c, we see that it takes larger values of $\beta$
to cause any change in the ground states for both AA and AB
configurations. In particular, for AA configuration at $\beta=0.1$,
$S_z=1$ remains the ground state for several $R$ values, unlike in
Fig. \ref{fig:Fig3}a. This is due to the screening of charged
impurities by electron-electron interactions. Also, $\beta_{cf}$ value
is increased to 0.36, consistent with single charged vacancy results
where $\beta_c$ is increased from 0.5 to 0.7 \cite{Polat2020}.

\section{\label{sec:level4}Summary}
To conclude, we have investigated the electronic and magnetic
properties of a system of two charged vacancies in hexagonal graphene
quantum dots using mean-field Hubbard approach. We focused on two
properties: (i) stability magnetic of phases of the two impurity
system and (ii) critical value of the Coulomb potential strength for
the frustrated collapse $\beta_{cf}$. We found that the magnetic
properties are sensitive to next nearest neighbor hopping parameter
$t_{nnn}$ and $\beta$. In particular, if $\beta$ approaches 0.2,
staggered magnetization is strongly suppressed pointing to a
non-magnetic regime within the subcritical region of molecular
collapse.  On the other hand, $\beta_{cf}$ is found to be nearly
constant for quantum dots sizes containing more than few thousands of
atoms. Finally, long range electron-electron interactions cause an increase up to 28\%  of $\beta_{cf}$ as a result of smearing out the electron
density near the Coulomb impurities.

\section{\label{sec:level5}ACKNOWLEDGMENT}
This research was supported by the Scientific and Technological
Research Council of Turkey T\"{U}B\.{I}TAK under the 1001 grant
project number 116F152, Turkey.
\newpage
\bibliographystyle{elsarticle-num}
\bibliography{references}

\end{document}